\documentstyle[preprint,aps,epsfig]{revtex}
\def\bea{\begin{eqnarray}}
\def\beann{\begin{eqnarray*}}
\def\beq{\begin{equation}}
\def\eea{\end{eqnarray}}
\def\eeann{\end{eqnarray*}}
\def\eeq{\end{equation}}
\def\nn{\nonumber}

\begin{document}
\draft
\title{The Gerasimov-Drell-Hearn Sum Rule and\\
the Single-Pion Photoproduction Multipole $E_{0+}$ close to 
Threshold} 
\author{D. Drechsel and G. Krein  
\thanks{\noindent
Permanent Address: Instituto de F\'{\i}sica Te\'{o}rica, \hfil\break
Universidade Estadual Paulista,
Rua Pamplona, 145\hfil\break
01405-900 S\~{a}o Paulo, SP - Brazil} 
}
\address{Institut f\"{u}r Kernphysik, Universit\"{a}t Mainz, 55099 
Mainz, Germany }

\maketitle
\begin{abstract}
The long-standing discrepancy between the Gerasimov-Drell-Hearn
sum rule and the analysis of pion photoproduction multipoles is greatly 
diminished by use of s-wave multipoles that are in accord with the
predictions of chiral perturbation theory and describe the experimental 
data in the threshold region. The remaining difference may be due to 
contributions of channels with more pions and/or heavier mesons whose 
contributions to the sum rule remain to be investigated by a direct 
measurement of the photoabsorption cross sections.
\end{abstract}


A great deal of our knowledge about the nucleon's ground state 
and its excited states has been obtained through experiments
with electromagnetic probes. The properties of the ground state can be 
related to photoabsorption cross sections through sum rules. The sum rule 
derived by
Gerasimov, Drell and Hearn (GDH)~\cite{gdh} is one of the most important 
ones; it provides an astounding relationship between the anomalous magnetic 
moment $\kappa$ of the nucleon and the photoabsorption cross sections for 
parallel and antiparallel alignments of the photon and photon helicities, 
$\sigma_{3/2}$ and $\sigma_{1/2}$, respectively. Specifically, the GDH sum 
rule is
\begin{equation}
- \frac{\kappa^2} {4} = \frac{M^2}{8 \pi^2 \alpha}
\int^{\infty}_{\omega_{th}} \frac{ \sigma_{1/2} (\omega) - \sigma_{3/2}
(\omega)} {\omega} \, d \omega ,
\label{defGDH}
\end{equation}
with $\omega_{th}$ the photoproduction threshold lab energy, 
$\alpha=e^2/4\pi=1/137$ the fine-structure constant, and $M$ the 
nucleon mass. The importance of this sum rule is due to the fact that it is
based on general principles of physics, such as Lorentz and gauge invariance, 
crossing symmetry, causality and unitarity. The sum rule has never been 
measured directly, but estimates for $\sigma_{1/2}$ and $\sigma_{3/2}$ have 
been made using pion photoproduction amplitudes~\cite{GDH-review}. The 
weighting factor $1/\omega$ in Eq.~(\ref{defGDH}) indicates that the 
low-energy region is very important for the sum rule. It is therefore to be
expected that a large fraction of the the sum rule is saturated by s-wave
near-theshold pion photoproduction and by $\Delta(1232)$ resonance production.

There exists an extensive literature of studies carried out in this 
direction~\cite{GDH-review}. Karliner's work~\cite{Karliner} is the first
one to include an estimate of the two-pion contribution to the sum rule, and 
the most recent studies are from Workman and Arndt~\cite{WA}, Burkert and 
Z. Li~\cite{H1:Bur93}, Sandorfi, Whisnant and Khandaker~\cite{SWK}, and 
Arndt, Strakovsky and Workman~\cite{ASW}. These analyses are usually 
performed by an isospin  decomposition of the 
photoproduction multipoles into isovector (VV), isoscalar (SS), and 
isovector-isoscalar (VS) components, so that the photoabsorption cross 
sections in  Eq.~(\ref{defGDH}) are given by 
$\sigma_{p,n}=\sigma^{VV}+\sigma^{SS} \pm \sigma^{VS}$ for protons 
and neutrons, respectively. Similarly we use, on the left hand side of 
Eq.~(\ref{defGDH}), the relations $\kappa_{p,n}=(\kappa^S \pm \kappa^V)/2$.
The general conclusions of these 
studies are that the $SS$ component is very small and the $VV$ component 
agrees reasonably well with the prediction of the sum rule, 
while there occurs an apparent discrepancy for the VS component, neither
its magnitude nor its sign agree with the sum rule. 
The solution for this discrepancy has usually been looked for in phenomena
occurring at the higher energies, e.g. in our
poor knowledge of two-pion photoproduction 
or a possible failure of convergence of the GDH sum rule ~\cite{Karliner}.

The purpose of the present communication is to draw attention to the 
somewhat unnoticed fact that the behavior of the $E_{0+}$ photoproduction 
multipole in the low energy region, close to the single-pion production
threshold is very important for this sum rule. In particular, we show that 
the use of an $E_{0+}$ amplitude that is in accord with low-energy theorems
and describes the experimental data diminishes considerably the discrepancies 
mentioned above.

The largest and most complete data base of photoproduction observables is
provided by the VPI-SAID program~\cite{SAID}. Although there have been
changes in the multipoles during the last years, mainly due to new
experimental data and reexaminations of
errors of older experiments, little has changed with respect to $E_{0+}$.
Very recently, Hanstein, Drechsel and Tiator (HDT)~\cite{HDT1}-~\cite{HDT3} 
have analyzed pion photoproduction imposing constraints from fixed-$t$
dispersion relations and unitarity. In the HDT approach, there are ten free 
parameters that are fitted to selected photoproduction data for photon
energies in the range of $160-420$~MeV. 
In particular, this data set contains
the new data from MAMI for differential cross sections of $\pi^0$
photoproduction off the proton near threshold~\cite{pi0}, and differential 
cross sections and beam asymmetries for $\pi^+$ and $\pi^0$ off the
proton~\cite{pi0pi+}. An interesting aspect of this approach is that 
the threshold region is not included in the data basis. Therefore, the 
threshold values obtained for the $s$-wave amplitudes are genuine predictions,
in the sense that the cross sections above $160$~MeV determine the threshold 
values by analytic continuation of the dispersion integrals. These predictions
are in excellent agreement with the results of chiral perturbation 
theory~\cite{BKM}. At threshold, the value of the amplitude 
$E_{0+}(n\pi^+)$ is $ 24.9 \times 10^{-3}/m_{\pi^+}$ in the SAID analysis
(version SP97K) and $28.4 \times 10^{-3}/m_{\pi^+}$ for HDT, $ 28.4 \times
10^{-3}/m_{\pi^+}$ predicted by ChPT~\cite{BKM} and 
$28.3 \pm 0.2 \times 10^{-3}/m_{\pi^+}$ according to the evaluation of
an older experiment~\cite{LI}. On the other hand, as has been clearly stated 
in Ref.~\cite{ASW}, the analysis in the very-low energy region becomes very 
complicated because of the different thresholds for $\pi^0 p$ and $\pi^+ n$
production and therefore the SAID multipoles should not be used in 
the $\pi^+ n$ threshold region.

The multipole decomposition of the numerator of the integrand of the GDH sum 
rule is~\cite{GDH-review}
\bea
\Delta \sigma &\equiv& \sigma_{1/2} - \sigma_{3/2} \nn\\
&=& 8 {\pi} \frac{q}{k} \sum_{l\ge 0}\frac{l+1}{2}\Bigl[
(l+2)\left(|E_{l+}|^2 + |M_{(l+1)-}|^2\right) \nn\\
&& - l \left(|M_{l+}|^2 + |E_{(l+1)-}|^2\right)\nn\\
&& + 2l(l+2)\left(E^*_{l+}M_{l+} - E^*_{(l+1)-}M_{(l+1)-}\right)
\Bigr] \nn\\
&=& 8 {\pi} \frac{q}{k} \Bigl(|E_{0+}|^2 + 3|E_{1+}|^2  + 
6 E^*_{1+}M_{1+}-|M_{1+}|^2 \nn\\
&& + |M_{1-}|^2 + \cdots \Bigr) ,
\label{intGDH}
\eea
where $q$ and $k$ are the c.m. momenta of the pion and the photon,
respectively. Note that $\Delta \sigma$ corresponds to -$2\,\sigma_{TT'}$ of
Ref.~\cite{GDH-review}.

The HDT analysis is limited to photon energies up to $500$~MeV, $s$,
$p$, and $d$ waves for isospin 1/2,  and $s$ and $p$ waves for isospin
3/2. For energies above $400$~MeV, the differences between the SAID 
and HDT multipoles are very small. However, large differences occur for the 
$E_{0+}$ multipole for $\pi^+$ production close to threshold, together with 
some minor differences for $M_{1+}$ below $300$~MeV. In Fig.~1 we present 
the comparison of both analyses for the integrand of Eq.~(\ref{defGDH}) 
up to $500$~MeV for the proton. More specifically, in Fig.~1(a) we plot 
the contribution of $E_{0^+}$, given by 
$8\pi\frac{q}{\omega k}\,|E_{0+}|^2$, and in Fig.~1(b) we plot
the contribution of $M_{1^+}$, $-8\pi\frac{q}{\omega k}\,|M_{1+}|^2$. 
In Fig.~1(c) we plot the sum of all multipoles to the integrand. 
As may be seen from Fig.~1(a), the HDT value for the $E_{0^+}$ contribution 
is substantially larger than in the case of SAID, in accordance with the 
threshold behavior of this amplitude as discussed above. Together with a much 
smaller (but opposite) effect for the $M_{1^+}$ multipole (see Fig.~1(b)), 
this clearly leads to a larger integrand in the case of HDT as shown in 
Fig.~1(c).

As a result the observed difference in these two multipoles the 
value of the integral of Eq.~(\ref{defGDH}) for the proton,
\beq
I_p = \int^{\infty}_{\omega_{th}} \frac{ \sigma^p_{1/2} (\omega) - 
\sigma^p_{3/2}(\omega)} {\omega} \, d \omega ,
\label{Ip}
\eeq
is changed by $20 \, \mu b$. Using the estimate of Karliner~\cite{Karliner} for
the two-pion contribution, $I_p(2\pi) = - 65 \, \mu b$ and the SAID multipoles
(SP97K solution) for the single-pion production, which gives 
$I_p(1\pi) = - 216 \, \mu b$, one obtains $I_p = - 281\, \mu b$ 
(Sandorfi et al.~\cite{SWK} obtain $I_p = - 289\, \mu b$ using another 
solution of the SAID multipoles).  This has to be compared to the GDH value,
$I_p = -281 \, \mu b$. Correcting then the one-pion contribution for the 
proper low energy behavior, we predict $I_p (1\pi) = - 196 \, \mu b$ and
$I_p = - 261 \, \mu b$, if we include the two-pion contribution as estimated
by Karliner. Expressed in different words, the discrepancy reduces from 38 \%
to 28 \% by use of s-wave multipoles that are in accord with the
low energy theorems and describe the experimental data in the threshold 
region. Concerning the remaining discrepancy it has to be said the estimate 
of the two-pion contribution of Ref.~\cite{Karliner} relies heavily on the 
assumption that the two-pion contribution is generated by the resonances, 
and that its helicity structure follows the known behavior of the one-pion 
contribution as given by Eq.~\ref{intGDH}. It is not obvious, however, that 
the two-pion background has to be resonance dominated, and it will be most 
interesting to see the outcome of the GDH experiment scheduled at 
MAMI and ELSA~\cite{MAMI}. For the neutron, the difference in 
the multipoles leads to a change by $17 \, \mu b$. This is a substantial
improvement, but still not enough to reverse the sign of $I^{VS}$. 

In Table~I we present the results of different studies of the sum rule. 
With the exception of the values given by Burkert and Li~\cite{H1:Bur93}, 
all results in the Table include the estimate of Karliner~\cite{Karliner} for 
the two-pion background. The results of the most recent analysis of Arndt, 
Strakovsky and Workman~\cite{ASW} are not presented in the Table because the 
authors do not quote their numbers, but do mention that their results are not 
very different from the ones of Ref.~\cite{SWK}. It is interesting to notice 
that if one uses the estimate of Burkert and Li for the contributions beyond
one-pion production, $-32\, \mu b$, together with the one-pion results of the 
HDT multipoles, the results are still closer to the prediction of the sum rule.
The discrepancy in this case would fall to 12 \% .

In conclusion, we would like to draw attention to the somewhat unnoticed fact 
that a precise threshold of the $E_{0+}$ single-pion photoproduction multipole
is quite essential for the GDH sum rule. The remaining discrepancies might be 
due to the non-resonant backgrounds.

\acknowledgments

Work partially supported by the Alexander von Humboldt Foundation and the
Deutsche Forschungsgemeinschaft SBF 201 (Germany) and FAPESP (Brazil).

\newpage

\onecolumn

\begin{figure}[h]
\centerline{
\epsfxsize=14.0cm
\epsfbox{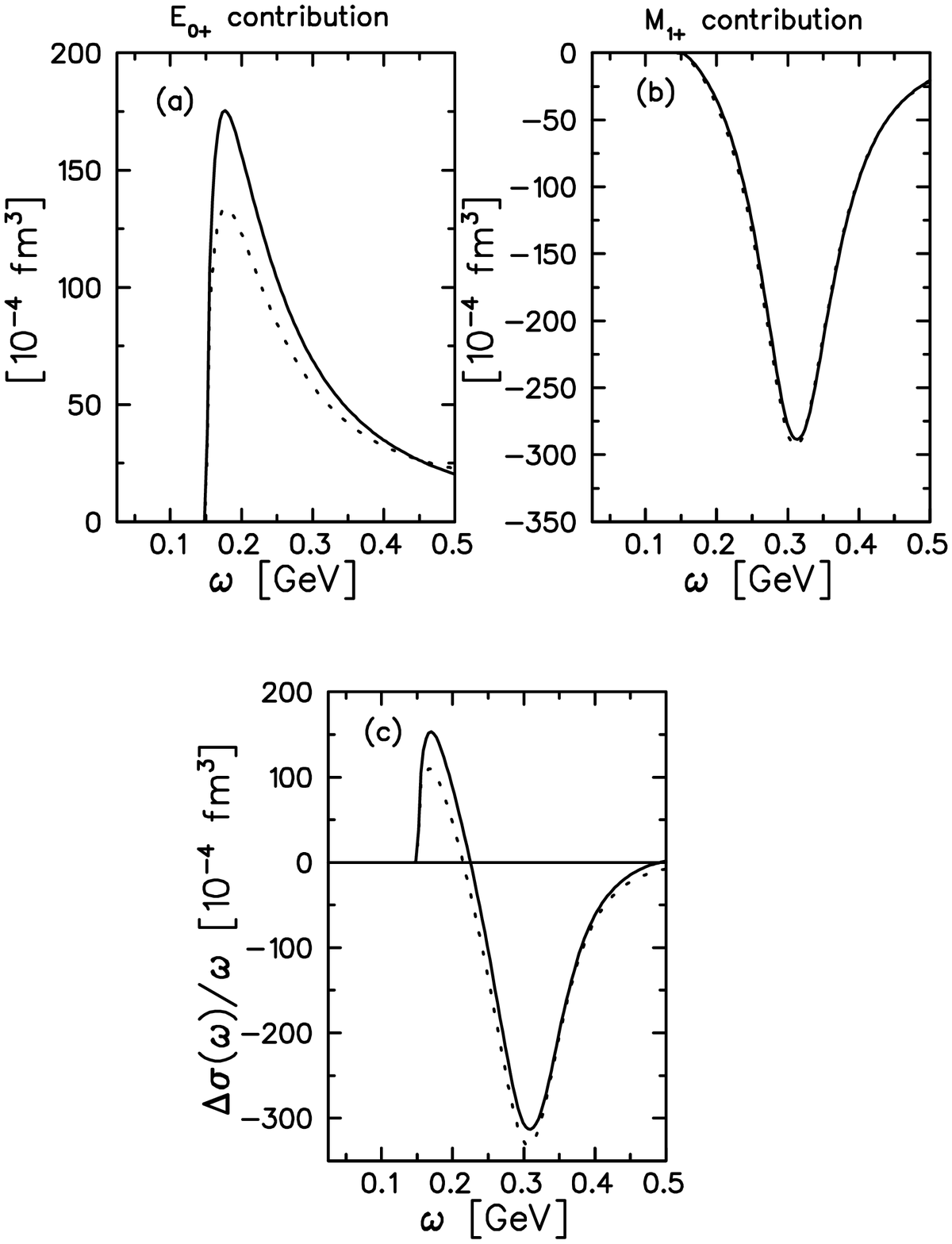}
}
\caption{
(a) Contribution of the multipole $E_{0^+}$ to the integrand of Eq.~(3), 
$8\pi\frac{q}{\omega k}\,|E_{0+}|^2$, (b) the corresponding contribution of
$M_{1^+}$, $-8\pi\frac{q}{\omega k}\,|M_{1+}|^2$, and  (c) the 
integrand for the complete calculation including all partial waves. The
solid lines correspond to the HDT multipoles and the dotted to the
SAID multipoles.
}
\end{figure}

\newpage

\begin{table}
\caption{Predictions from various
models and data analyses for the GDH integral for proton $(I_p)$, 
neutron $(I_n)$, and the difference $I_p - I_n$ in units of $\mu b$. 
The results, with exception of the ones of Ref.~[5], include the two-pion 
background as estimated by Karliner~[3]. }
\begin{center}
\begin{tabular}{l|r|r|r}
                                &$I_p$   &$I_n$     &$I_p-I_n$  \\
\hline
GDH integral                    &-204.5  &-232.8    &28.3       \\
\hline
Karliner~\cite{Karliner}        &-261    &-183      &- 78       \\
Workman and Arndt~\cite{WA}     &-260    &-157      &-103       \\
Burkert and Li~\cite{H1:Bur93}  &-203    & --       & --        \\
Sandorfi et al.~\cite{SWK}      &-289    &-160      &-129       \\
This work                       &-261    &-180      &- 81
\end{tabular}
\label{H1:GDH}
\end{center}
\end{table}

\end{document}